%% file: main.tex
\documentclass{WileyMSP-template}

\usepackage[super, numbers,sort&compress]{natbib}
\usepackage{hyperref}
\usepackage[utf8]{inputenc}
\usepackage{amsmath}
\usepackage{amssymb}
\usepackage{siunitx}
\usepackage{xcolor}
\usepackage{lineno}
\begin{document}

\pagestyle{fancy}
\rhead{\includegraphics[width=2.5cm]{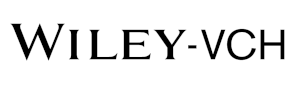}}

\title{Toggling stiffness via multistability}

\maketitle

% Author: Please give full first and last names for authors and include * after the name of all corresponding authors

\author{Hugo de Souza Oliveira*,}
\author{Renate Sachse*,}
\author{Michele Curatolo,}
\author{Edoardo Milana}
% Dedication

\dedication{*Hugo de Souza Oliveira and Renate Sachse contributed equally to this work.}

\begin{affiliations}
H.S. Oliveira and E. Milana\\
Department of Microsystems Engineering - IMTEK,\\ University of Freiburg, Freiburg, Germany\\
Cluster of Excellence livMatS @ FIT—Freiburg Center for Interactive Materials and Bioinspired Technologies,\\ University of Freiburg, Freiburg, Germany\\
Email Address: edoardo.milana@imtek.uni-freiburg.de

M. Curatolo\\
Architecture Department,\\ University of Roma Tre, Rome, Italy\\

R. Sachse\\
Max Planck Institute for Intelligent Systems, Stuttgart, Germany\\
John A. Paulson School of Engineering and Applied Sciences,\\
Harvard University

\end{affiliations}

% Keywords: Please provide a minimum of three and a maximum of seven keywords, separated by commas


% Abstract should be written in the present tense and impersonal style (i.e., avoid we), and be at most 200 words long
\begin{abstract}
Variable stiffness is a key capability in biological and robotic systems, enabling adaptive interaction across tasks and environments. Mechanical metamaterials offer an alternative to conventional mechatronic solutions by encoding stiffness variation directly into monolithic structural architectures, reducing the need for discrete assemblies. Here, we introduce a multistable mechanical metamaterial that exhibits a toggleable stiffness effect in which the effective shear stiffness switches discretely between stable mechanical configurations. Mechanical analysis of surrogate beam models of the unit cell reveals that this behavior originates from the rotation transmitted by the support beams to the curved beam, governing the balance between bending and axial deformation. Consequently, the shear stiffness ratio between the two states can be tuned by varying the slenderness of the support beams or by incorporating localized hinges that modulate rotational transfer. Experiments on 3D-printed prototypes validate the numerical predictions and confirm consistent stiffness toggling across different geometries. Finally, we demonstrate a monolithic soft clutch that leverages this effect to achieve programmable, stepwise stiffness modulation. This work establishes a design strategy for toggleable stiffness using multistable metamaterials, with potential applications in soft robotics and smart structures where adaptive compliance is of paramount importance.\end{abstract}

% Text: Please use section headings and subheadings as specified below. For communications, all section headings apart from Experimental Section should be removed
% Please make the first reference to a display item bold: \textbf{Figure 1}
% Do not abbreviate Figure, Equation, etc.; display items are always singular, i.e., Figure 1 and 2.
% Equations are always singular, i.e., Equation 1 and 2, and should be inserted using the {equation} environment, not as graphics
% Please do not use footnotes in the text, additional information can be added to the Reference list.

\input{1_introduction}

\input{2_results}

\input{4_conclusions}
\input{3_methods}

\medskip
%\textbf{Supporting Information} \par %Please delete the Suppporting Information statement if it is not applicable. Please supply Supporting Information in another file. Supporting information should not be provided in .tex format
% Supporting Information is available from the Wiley Online Library or from the author.

% Acknowledgements
\medskip
\textbf{Acknowledgements} \par %delete if not applicable))
This research is funded by the Deutsche Forschungsgemeinschaft (DFG, German Research Foundation) under Germany’s Excellence Strategy – EXC-2193/1 – 390951807.

% References
\medskip

\bibliographystyle{MSP}
\bibliography{references}

\end{document}

%% file: 1_introduction.tex
\section{Introduction}

Metamaterials derive their unconventional physical properties from the deliberate architectural arrangement of material and geometry. When such engineered responses manifest in stiffness, Poisson’s ratio, or kinematics, they are referred to as mechanical or architected metamaterials \cite{bertoldi_flexible_2017,jiao2023mechanical,surjadi2019mechanical,de_souza_oliveira_mechanical_2025}. By encoding functionality directly into structure, mechanical metamaterials enable behaviors that transcend those of their constituent materials.

Recent efforts have explored how architected materials can contribute to autonomous material systems, in which control and functionality are embedded at the physical level rather than relying on external electronics \cite{Aubin2024,de_souza_oliveira_mechanical_2025}. Such concepts are being applied in soft robotics, where low-level controllers are implemented in the nonlinear physical properties of soft devices \cite{milana_physical_2025}. Such physically embodied control has enabled oscillations \cite{rothemund_soft_2018, van_laake_fluidic_2022, comoretto_physical_2025}, programmed sequences \cite{milana_morphological_2022,van_raemdonck_nonlinear_2023}, and reactive behaviors \cite{drotman_electronics-free_2021, comoretto_embodying_2025}. These capabilities are rooted in strongly nonlinear mechanics, particularly snap-through instabilities and multistability \cite{pal_exploiting_2021}. As such, mechanical metamaterials can offer many possibilities, due to the large design spaces and fine tuning of nonlinear mechanical properties.

Metamaterials can also be used to emulate functionalities from biological systems. For example, the ability to modulate mechanical stiffness. Variable stiffness enables adaptive interaction with the environment across spatial and temporal scales. For example, wind-exposed trees adjust branch stiffness during growth to withstand sustained loads \cite{bruchert2006effect}, while sea cucumbers reversibly alter body stiffness through biochemical triggers \cite{thurmond1996morphology}. Humans continuously regulate grasp stiffness depending on task requirements \cite{napier_prehensile_1956}.  Variable stiffness is of paramount importance in robotics as well, to augment physical capabilities in manipulation and locomotion. In robotics, stiffness modulation is typically achieved using mechatronic assemblies such as Series Elastic Actuators (SEAs) and Variable Stiffness Actuators (VSAs) \cite{wolf_variable_2016}. While effective, these systems introduce additional mechanical complexity, mass, and inertia, which can limit performance and partially offset the benefits of adaptive compliance \cite{vu_improving_2016,khemakhem_impact_2025}. 

In the context of variable stiffness, mechanical metamaterials offer an alternative pathway, where stiffness variation emerges from monolithic structural architecture rather than from discrete components. In particular, multistable mechanical metamaterials, which are systems with multiple local minima in their elastic energy landscape, provide a route toward discrete stiffness modulation. Bistable architectures can encode binary stiffness along linear or rotational motion \cite{kuppens2021monolithic}, and origami-inspired designs such as Kresling structures exhibit distinct flexural stiffness in different stable states \cite{kaufmann}.  Because the stiffness change is not continuous like in conventional VSAs but instead switches discretely between mechanically stable configurations, we refer to this phenomenon as \textit{toggleable stiffness}.

% ------------------------------- HUGO----------

\begin{figure}[!b]
    \begin{center}
        \includegraphics[width=1.0\textwidth]{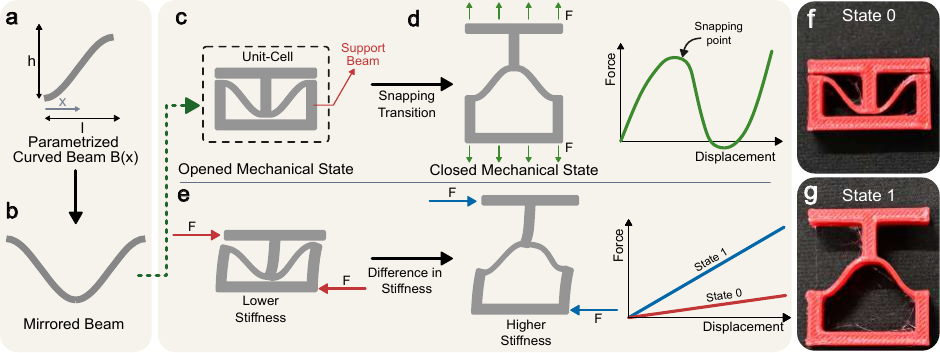}
    \end{center}
    \caption{Conceptual overview of the bistable unit cell: (a) parametrized curved beam $B(s)$; (b) mirrored beam; (c) unit cell geometry with support beam in the opened mechanical state; (d) snapping transition to the closed mechanical state and corresponding conceptual force--displacement response; (e) shear-loading condition showing the difference in stiffness between State~0 and State~1; (f) 3D-printed bistable unit cell in State~0; (g) 3D-printed bistable unit cell in State~1.}    
    \label{intro}
\end{figure}

Here, we reveal toggleable shear stiffness in multistable mechanical metamaterials undergoing snap-through under tensile loading. Such systems have been extensively investigated, primarily focusing on their tensile snapping behavior across variations in beam geometry, thickness, length, material composition, and three-dimensional topology \cite{yang_1d_2020,shan_multistable_2015,restrepo_phase_2015,rafsanjani_snapping_2015}. These metamaterials consist of arrays of repeating unit cells formed by two mirrored beams that exhibit snap-through behavior analogous to a von Mises truss \cite{mises_uber_1923} (Figures~\ref{intro}a--d). We show that these unit cells exhibit distinct shear stiffness in each of their stable mechanical states (Figures~\ref{intro}e), with the configuration with larger height exhibiting higher effective shear stiffness. We elucidate the mechanical origin of this configuration-dependent stiffness and construct a finite-element-based design map for the planar topology, which is experimentally validated using 3D-printed metamaterials. Finally, we demonstrate the functional potential of this mechanism through a monolithic soft clutch that translates shear-induced stiffness modulation into tunable uniaxial rigidity.

%% file: 2_results.tex
\section{Results}

\subsection{Shear deformation of the bistable unit cell}

Mechanical metamaterials that snap under tension are typically composed of a unit cell which features a curved beam shape, such as the Euler first buckling mode \cite{yang_multi-stable_2019}. Similarly, in our case we explore the curve $B(x)$ (Figure~\ref{intro}a), parametrized as:

\begin{equation}
    B(x) = \left(\frac{h}{2}\right)\left(1-\cos\left(2\pi\left(\frac{x}{l}\right)\right)\right). 
\end{equation}

This curved beam is attached to the frame via three vertical support beams (Figure~\ref{intro}c). The global snap-through effect of the metamaterial emerges from the local snap-through of each unit cells. When a normal force is applied to a unit cell, the force-displacement displacement response exhibits a nonmonotonic behavior. Upon intersecting the zero-force axis, a second stable state is reached, as shown in Figure~\ref{intro}d. This effect is well reported in the literature about snapping metamaterials \cite{yan_wu_wen_sun_zhou_2025}.

However, to the best of our knowledge, no studies report on the mechanical characteristics of these metamaterials, when a shear force is applied. Interestingly when the unit cell is in its primary stable state (State 0 from now on in the manuscript, Figure~\ref{intro}e) and subjected to shear loading, the response is linear with a constant structural shear stiffness. In contrast, when the unit cell is snapped into the secondary state (State 1, Figure~\ref{intro}f) and the same shear loading is applied, the structure responds with a higher structural shear stiffness.   

This difference in shear stiffness strongly depends on the geometry of the support beams. When subjected to shear loading, the support beams undergo bending, causing a rotation at the attachment points of the curved beam. Consequently, not only the horizontal force is transmitted to the curved beam, but also a moment. As we will demonstrate in the next section, the different mechanical behavior of the two states is due to the deformation induced by this moment.

\subsection{Mechanical analysis of the unit cell}
\label{mec-analysis}

To evaluate the stiffness ratio between the two states, we develop two simplified finite-element models of the unit cell based on beam elements. We define the shear stiffness ratio as $ \eta = \frac{k_0}{k_1}$, where $k_0$ and $k_1$ define the shear stiffness of the States 0 and 1, respectively. From this point onward in the manuscript, structural shear stiffness will be referred to simply as stiffness to avoid unnecessary repetition and maintain readability.

\begin{figure}
    \centering{\tiny
\includegraphics[width=1.0\textwidth]{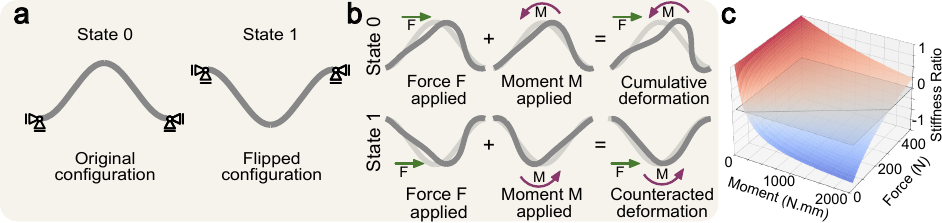}}
    \caption{Model A: (a) State 0 and State 1 configurations; (b) Deformations for State 0 and State 1 under isolated horizontal force $F$, isolated bending moment $M$, and the combined load case. (c) Parameter sweep showing the shear stiffness ratio between states as a function of the applied bending moment $M$.}
\label{fig:simplified_model_deformation}
\end{figure}

The first model (Model A) only considers the curved beam pinned at both ends (Fig.~\ref{fig:simplified_model_deformation}a). A horizontal force $F$, and a moment $M$ are applied to the center of the structure, which is the connecting point with the support beam, as discussed above. The moment $M$ is applied to induce rotation, simulating the effect of the bent supported beams.  Because of the geometry of the structure the rotation also generates a horizontal displacement. Additionally, the horizontal force $F$ causes an horizontal displacement as well, the one induced by the shear load. For simplicity, in this model we neglect the effect of vertical displacement. 

The difference between the two states is introduced by flipping the direction of the moment, under the ideal assumption that in State 1 the curved beam goes into a perfectly mirrored configuration (Figure~\ref{fig:simplified_model_deformation}a). Figure~\ref{fig:simplified_model_deformation}b illustrates the resulting deformation patterns for both stable states under the different load cases. In State 0, the horizontal displacements induced by $F$ and $M$ are aligned. Consequently, when both loads are applied simultaneously, their effects accumulate, resulting in larger deformations. However, in State 1, the rotation is in opposite direction, causing a counteracting effect when the load cases are combined, making the structure appear stiffer, as the same load combination results in a smaller horizontal deformation.

A parameter sweep of the bending moment $M$ and the horizontal force $F$(Fig. \ref{fig:simplified_model_deformation}c) shows that the stiffness ratio between the two states is highly sensitive to the amount of rotation transferred to the beam. Although this analysis is an abstraction, as in reality the moment and force are not independent, it helps us to validate our hypothesis. The results in the limiting case of $M \rightarrow 0$ show that the stiffness ratio approaches unity, as expected since the load in the two cases is identical. As the bending moment $M$ (and thus the rotation) increases, the stiffness ratio decreases, meaning that the difference between the two states becomes more pronounced. These results establish the rotation of the curved beam\textquotesingle s center as the governing parameter of the stiffness ratio.

\begin{figure}[!t]
    \centering
    \includegraphics[width=1\linewidth]{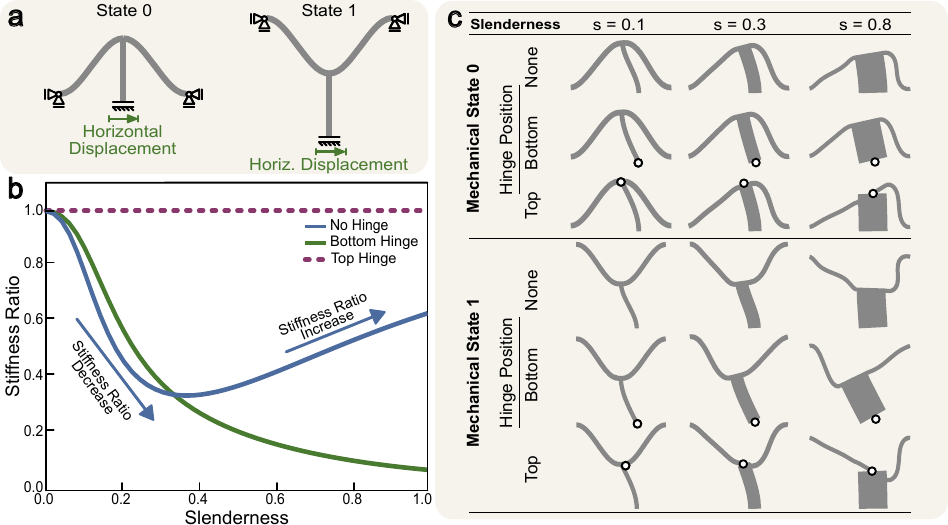}
    \caption{Model B: (a) Model of the unit cell in the beam analysis. b) Variation of the stiffness ratio with the slenderness of the support beams in the case of no hinge, bottom hinge and top hinge. c) Deformation patterns for State 0 and State 1 at representative inverse slenderness ratios ($s = 0.1$, $s = 0.3$, $s = 0.8$) for all hinge configurations.}
    \label{modelB}
\end{figure}

Further, we introduce a second model (Model B, Fig.~\ref{modelB}a) that includes the support beams, providing a more accurate analysis of the unit cell. The structure is loaded by imposing a horizontal displacement at the lower end of the support beam. In this configuration, the coupling between force and rotation transmitted to the curved beam is primarily governed by the slenderness of the support beam, making it a key parameter to control the stiffness ratio in the design, as illustrated in Fig.~\ref{modelB}b. 

Fig.~\ref{modelB}c shows deformation patterns for three representative cases of slenderness. For practical reason, we use an inverse slenderness ratio  $s$ ($s = \frac{c}{L}$, width over length ratio). The three cases represent high slenderness ($s=0.1$), intermediate ($s=0.3$), and low ($s=0.8$). For high slenderness $s < 0.2$, the support beam is highly compliant in bending, most of the  displacement is accommodated locally within the support itself. As a result, only limited rotation and force are transmitted to the curved beam. Consequently, the overall stiffness ratio between the two states decreases. Conversely, for low slenderness (thick support beams), high horizontal force but very little rotation is transmitted. Thus, an overly stiff beam also reduces the stiffness ratio. In the extreme case in which the support beam would be completely rigid, there would be no bending moment to transmit. According to Figure~\ref{modelB}, an optimal inverse slenderness ratio of approximately 0.3 minimizes the stiffness ratio down to 0.35 (note that $k_1>k_0$, so $0<\eta\le1$) . At this point, the beam balances force transfer and rotation, resulting in the strongest toggleable stiffness effect.

To modulate the toggleable stiffness effect, we can also directly control the rotational transfer at the interface between the support and curved beams. We introduce two scenarios in our Model B by including hinges at the extremes of the support beam (Fig.~\ref{modelB}). In the first scenario, a hinge is placed at the junction between the support and curved beams. This configuration effectively eliminates any moment transfer into the curved beam. In this case, no bending moment is introduced, and the deformation of the curved beam is identical for the two states, leading to a stiffness ratio of unity, as shown in Fig.~\ref{modelB}b. In the second scenario, the hinge is positioned at the intersection between the support beam and the frame (bottom hinge). This arrangement reduces the effect of horizontal force but amplifies the rotations for low slenderness, effectively generating very low stiffness ratios. Details on the simplified beam models (Model A and Model B) are reported in the Supplementary Information.

\subsection{Two-cell mechanical metamaterial}

\begin{figure}[!b]
    \centering
    \includegraphics[width=1\linewidth]{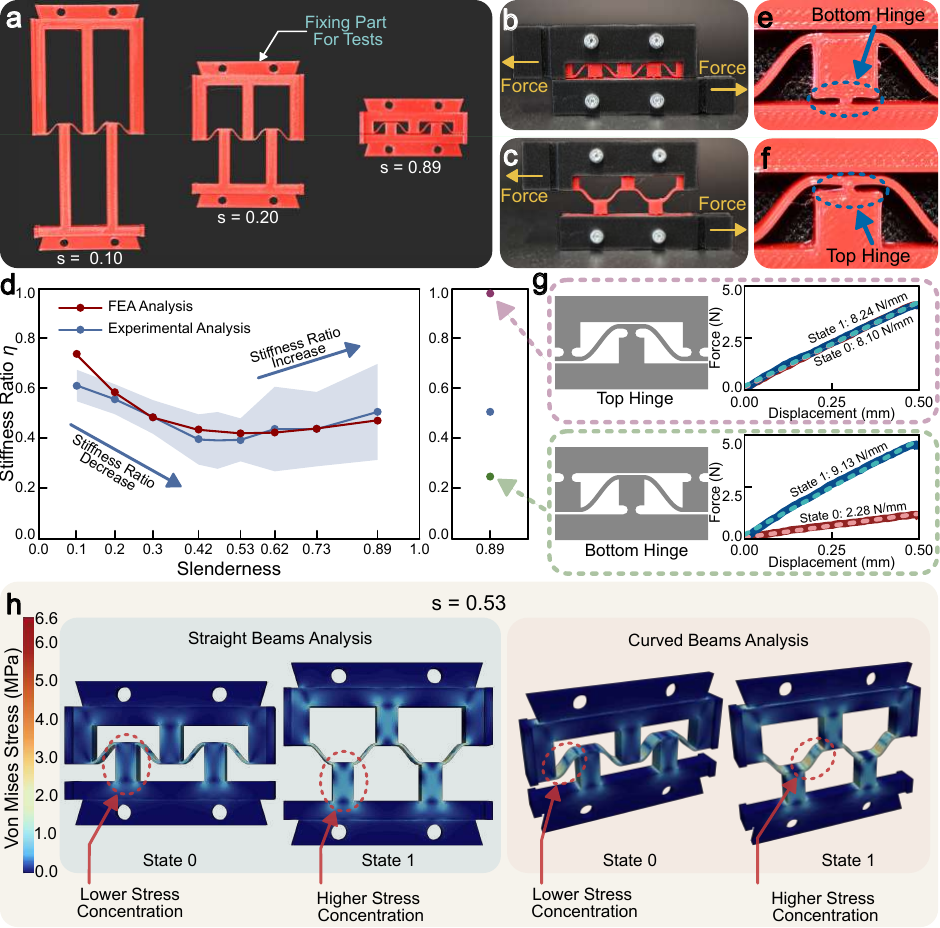}
    \caption{Mechanical characterization of two-cell metamaterials: (a) fabricated specimens with varying inverse slenderness $s$; (b) experimental setup showing the fixation system for shear testing in State 0; (c) and State 1; (d) stiffness ratio $\eta$ as a function of inverse slenderness $s$, obtained from experiments and finite element analysis; (e) bottom-hinge configuration with notch flexure; (f) top-hinge configuration with notch flexure; (g) force-displacement curves for top-hinge and bottom-hinge configurations; (h) von Mises stress distribution for State 0 and State 1 at $s = 0.53$ from finite element simulations.}
    \label{Hinges}
\end{figure}

Building upon the unit cell mechanical analysis, we design a planar bistable metamaterial consisting of two adjacent cells to experimentally characterize the toggleable stiffness effect. The structures are fabricated via fused deposition modeling (FDM) using thermoplastic polyurethane (TPU) filament, as detailed in the Materials and Methods section. Shear characterization is performed using a universal testing machine. Each specimen is tested sequentially in State 0 (the printed configuration) and State 1 (snapped configuration), and the effective structural stiffness is extracted from the measured force–displacement response. Figure~\ref{Hinges}a presents representative structures with different support-beam inverse slenderness ratio $s$. A top-mounted fixing feature is incorporated to secure the specimens during shear testing, as illustrated in Figure~\ref{Hinges}b and c for State 0 and State 1, respectively. The general dimensions of the structure can be referred to in Figure S2a.

Additionally, the response of the metamaterial under shear is simulated using Finite Element Analysis (FEA), to corroborate the experimental data. Figure~\ref{Hinges}d presents the stiffness ratio $\eta$ obtained experimentally and via FEA under an imposed displacement of \SI{0.5}{\milli\metre}. In both cases, $\eta$ exhibits a non-monotonic dependence on the inverse slenderness $s$, reaching a minimum near $s \approx 0.53$. The lowest stiffness ratios are approximately $\eta \approx 0.40$ experimentally and $\eta \approx 0.43$ from FEA. For both smaller and larger values of $s$, $\eta$ increases, indicating the presence of a minimum, corresponding to an optimal configuration that maximizes the difference in stiffness between the two states, similar to the trend predicted by the mechanical analysis of the unit cell (Figure \ref{modelB}b). Moreover, the good agreement between experiments and 3D FEA proves that this effect is purely caused by the geometric reconfiguration of the unit cell, and there is no significant influence of the elastic energy stored in the metastable state (State 1). The experimental snapped configuration is well approximated by the exact mirrored geometry of the curved beam, which was used in the simulations of the unit cell as well as the 3D metamaterial.

We further experimentally investigate the influence of hinge placement on the metamaterial response. The hinge effect is realized by introducing notch flexures at the ends of the support beams with $s = 0.89$, since this effect is larger at lower slenderness according to our model (Figure \ref{modelB}b). The resulting bottom-hinge and top-hinge configurations are illustrated in Figure~\ref{Hinges}e and f, respectively. The hinges significantly affect the stiffness ratio, consistent with the predictions of Model B. As shown in Figure~\ref{Hinges}g, the top-hinge configuration suppresses rotational transfer and consequently prevents bending moment transmission to the curved beams. Under this condition, only axial force is transmitted in both State~0 and State~1, resulting in $\eta \approx 1$. On the contrary, in the bottom-hinge configuration the beam rotates around the flexure hinge instead of bending, therefore offering less resistance to imposed displacement.  As a result, the stiffness ratio reaches its minimum value, $\eta \approx 0.24$, indicating a pronounced stiffness contrast between the two states. Figure~\ref{Hinges}g also presents the corresponding force-displacement curves for both hinge configurations, demonstrating the stiffness difference introduced by hinge positioning.

Figure~\ref{Hinges}h illustrates the simulated von Mises stress distribution along the structure for a representative inverse slenderness of $s=0.53$ under identical shear loading conditions. A clear increase in stress concentration is observed in State 1 compared to State 0. This amplification originates only from geometric factors, as the simulations only consider purely mirrored configurations of the curved beams (no energy is initially stored in State 1). The elevated stress concentration in State 1 is therefore a direct manifestation of the enhanced local deformation resistance responsible for the higher effective stiffness measured experimentally and numerically.

\subsection{Monolithic soft clutch demonstrator}
\label{demonstrator}
\begin{figure}
    \centering
    \includegraphics[width=0.9\linewidth]{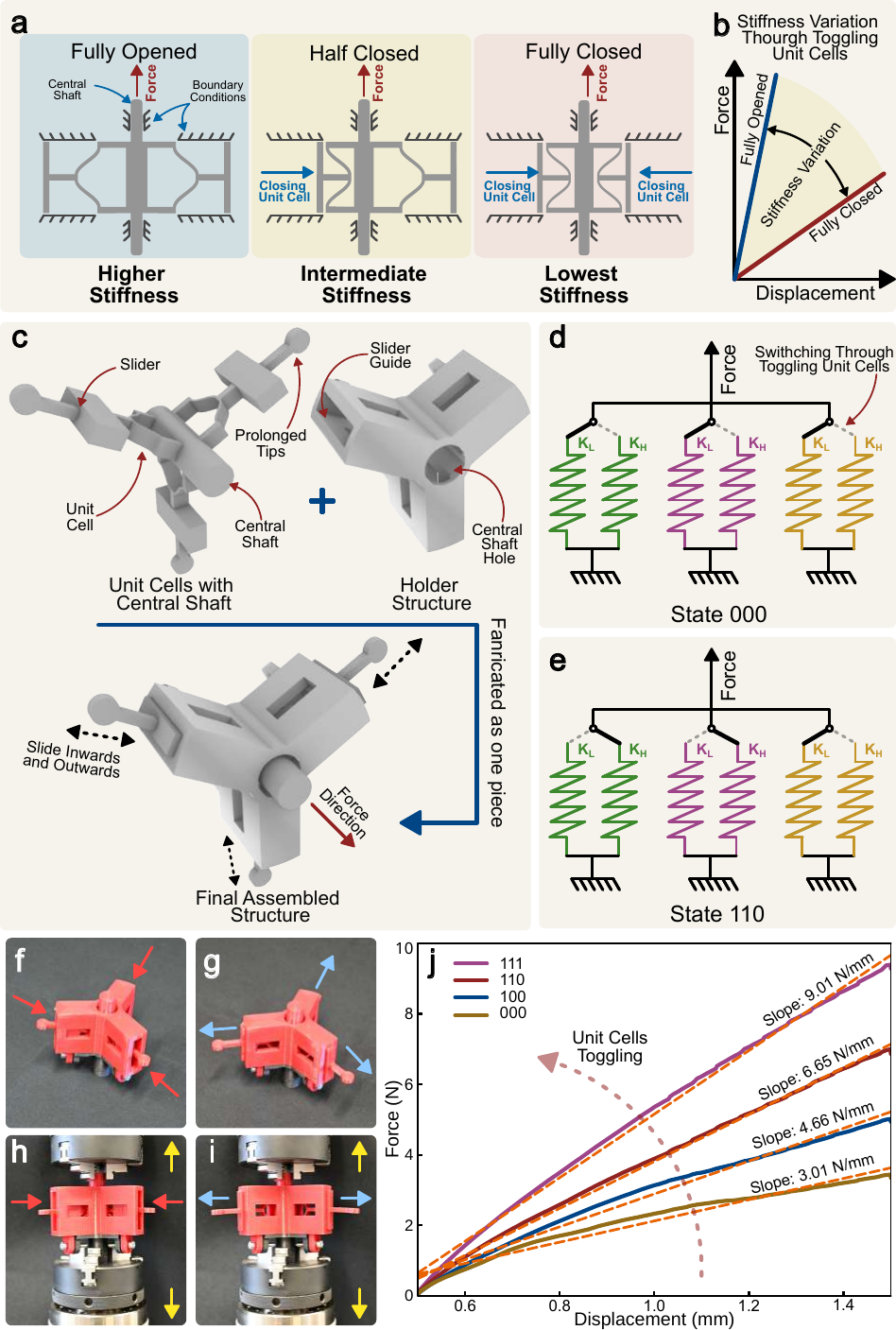}
    \caption{Monolithic soft clutch based on toggleable unit cells. (a) Conceptual representation of unit cells connected to a central shaft with sliding boundary conditions. (b) Schematic of stiffness variation as unit cells transition between closed ($k_0$) and open ($k_1$) states. (c–e) Design and assembly steps of the demonstrator. (f–i) Fabricated device showing unit cells toggled between the two states and tested under tensile loading. (j) Force–displacement curves for the four possible states (000, 100, 110, 111), showing incremental stiffness increase as more unit cells are in the open state.}
    \label{dem1}
\end{figure}

To showcase the application potential of the toggleable stiffness effect in soft machines, we developed a monolithic soft clutch as demonstrator. Indeed, these unit cells can serve as building blocks for more complex materials that operate based on the resulting shear stiffness. Fig.~\ref{dem1}a illustrates the conceptual principle of the soft clutch, which can toggle between multiple stiffness according to the number of bistable unit cells. For example, two unit cells can be connected in parallel to a central shaft and constrained by boundary conditions that allow only sliding motion. As a result, axial loading of the shaft induces shear deformation in the unit cells. Each unit cell switches between State 0 and State 1 associated to a low $k_0$ and a high $k_1$ stiffness, respectively, acting as shear springs. In Fig.~\ref{dem1}a, two unit cells State 1 (11) make the structure exhibit higher stiffness. Then, by toggling one unit cell, the overall stiffness is at an intermediate value (01). By toggling the other unit cell (00), the structure yields the lowest stiffness. The conceptual stiffness variation between fully opened and fully closed cases is illustrated in Fig.~\ref{dem1}b. This concept is applied to create a device where up to four different stiffness values can be achieved by toggling three unit cell, behaving as a clutch. The design of this monolithic soft clutch is shown in Fig.~\ref{dem1}c. Three unit cells having sliders and prolonged tips attached to their top parts are connected to a central shaft. Although we observed that adding notch flexures at the bottom of the support beams lead to largest increase in stiffness difference (Figure~\ref{Hinges}d), this design is unpractical due to overall reduction of stiffness as well as the weakness of the notch, prone to mechanical failure. Therefore, we use support beams with inverse slenderness $s=0.4$, in the range that minimizes the stiffness ratio (Figure~\ref{Hinges}d). This first structure is inserted inside a holder structure that represents the constraining boundary conditions illustrated in Fig.~\ref{dem1}a, resulting in the final mechanical device. Details on the dimensions are reported in the Supplemental Information in Figures S2b and c. The device is monolithic and soft because we print it as a single piece, using multi-material FDM 3D printing and TPU soft filaments.

Toggling the unit cells through the prolonged tips switches between the two mechanical states. Thus, only four distinct stiffness levels are possible, which are 000, 100, 110, and 111, where each digit denotes the state of the unit cell. Since the unit cells are designed using identical geometrical parameters, permutations of 0 and 1 yields the same mechanical response, regardless of their spatial arrangement. For example, when only one unit cell is in State 1, 100 results in the same behavior of 001. Figs.~\ref{dem1}d,e illustrate, for example, the cases when all unit cells are in State 0 (000) and when only one unit cell is in it (110), respectively. Effectively, the unit cells are in a parallel configuration, and the structural stiffness add up. Thus, the equivalent stiffness $k_{eq}$ of the structure can be calculated as in eq.~\eqref{eq-stiff}, in which $n$ represents the number of State 1 unit cells, $n = 0,1,2,3$. 

\begin{equation}
    k_{eq}(n) = (3-n)k_0 + n k_1
    \label{eq-stiff}
\end{equation}

Figs.~\ref{dem1}f,g show the structure fabricated as a monolithic piece with prolonged tips inwards and outwards, resulting in the unit cells being in State 0 and State 1, respectively. Figs.~\ref{dem1}h,i show the structure being held in a tensile test machine. Fig.~\ref{dem1}j shows the force-displacement curves obtained by toggling the mechanical states of each unit cell. For the state 000, the equivalent stiffness of the structure is \SI{3.01}{\newton\per\milli\meter}. The state 100 increases the equivalent stiffness by $\approx$ 1.55, reaching $k_{eq} = \SI{4.66}{\newton\per\milli\meter}$. Sequentially, the states 110 and 111 increase the equivalent stiffness by $\approx$ 1.42 and $\approx$ 1.35, reaching $k_{eq} = \SI{6.65}{\newton\per\milli\meter}$ and $k_{eq} = \SI{9.01}{\newton\per\milli\meter}$, respectively. Although the unit cells were designed with identical geometrical and fabrication parameters, the incremental stiffness gains are not perfectly uniform, as for the ideal summation. This deviation is likely due to the small imperfections in the fabrication through 3D printing process, as well as variation in the mechanical properties of the soft material.

%% file: 4_conclusions.tex
\section{Conclusions}

This work introduced and explained the \textit{toggleable stiffness} effect in multistable mechanical metamaterials that snap under tension and are loaded under shear. Through a combination of simplified beam models,  3D finite element structural simulations, and experiments on 3D-printed prototypes, we demonstrated that the stiffness difference between stable states originates from purely geometrical factors, which affect the rotation transmitted to the curved beams by the support beams. The rotation transmission is the decisive factor in controlling the stiffness ratio. The amount of this rotation, and thus the magnitude of the stiffness ratio, is primarily governed by the slenderness of the support beams, which controls the balance between bending compliance and rotational transfer. Thus, the slenderness of the support beam effectively acts as a design parameter, allowing the mechanical response to be tuned between different compliant configurations. Moreover, introducing localized compliance through flexures can further extend the design space, providing fine control over how rotation and force are coupled. 

Building upon these mechanical insights, we developed a monolithic soft clutch demonstrator that embodies the toggleable stiffness principle. By arranging bistable unit cells in parallel, the device achieves discrete, stepwise stiffness modulation, with the number of toggled unit cells determining the overall structural stiffness. This demonstrator highlights the potential of toggleable stiffness metamaterials to create in future passive or semi-active soft mechanical devices, capable of adapting their rigidity without external actuators or complex control systems, offering a route to applications in adaptive robotic manipulation and locomotion.

Overall, this work provides both a physical understanding and a design methodology for creating multistable metamaterials with toggleable stiffness modulation. Such materials open new avenues for soft robotic systems, adaptive structures, and vibration control devices, where stiffness switching can be directly embedded at the (meta)material level. Future work will explore physical control to switch between stiffness states through the integration with soft actuators for autonomous toggling, and translating the concept to smaller scales.

%% file: 3_methods.tex
\section{Materials and methods}

\subsection{Mechanical analysis of the unit cell}
\label{Mechanical_Analysis}

Finite element simulations are performed in Abaqus (Abaqus/CAE 2021, Dassault Systèmes) to characterize the deformation behavior of the bistable unit cell in its two stable states under shear. All simulations were conducted in SI units (mm, N, MPa, s). Both models (Model A and B discussed in the Results section) are discretized using two-dimensional planar, deformable wire elements (B21 element type, 2-node linear beam in a plane). This element is based on Timoshenko beam theory, allowing for transverse shear deformation and can be used for thick as well as slender beams. The global mesh size is set to \SI{0.5}{\milli\metre}, ensuring sufficient resolution. We investigate two primary model configurations:

\begin{itemize}
    \item Model A: Two curved beams (Figure S1a,b): Forces ($F$) and moments ($M$) are applied at the center.
    \item Model B: Two curved beams connected to a support beam (Figure S1a,c): A horizontal displacement of \SI{5}{\milli\metre} is applied to the lower end of the support beam.
\end{itemize}

The material is modeled as linear elastic with a Young’s modulus of \SI{1000}{\mega\pascal} and a Poisson’s ratio of 0.4, focusing the analysis on geometric nonlinearities. Periodic boundary conditions are applied to represent the behavior of an infinite array, constraining translations at the beam ends (u1 = u2 = 0) and coupling their rotations (ur3) to a central control point to simulate neighboring cell interactions.
In Model B, the support beam is rigidly connected to the curved beams at their intersection midpoint, with its lower end constrained in vertical displacement (u2 = 0) and rotation (ur3 = 0) while allowing horizontal movement. For Model A, concentrated forces ranging from \SI{0}{\newton} to \SI{500}{\newton} and moments ranging from \SI{0}{\newton\milli\metre} to \SI{2000}{\newton\milli\metre} are applied at the center of the curved beams to the control point that couples the beam-end rotations. For Model B, a horizontal displacement of \SI{5.0}{\milli\metre} is applied to the lower end of the vertical support beam.  In certain model variants, hinges are introduced to modify connection behavior: a top hinge allowed relative rotation by coupling only displacements (u1, u2) between the curved and support beams, whereas a bottom hinge released the rotational constraint (ur3) at the support while maintaining vertical fixation (u2 = 0).

\subsection{FEA simulations of the metamaterials}
Numerical simulations of the two-cell metamaterials are performed using the COMSOL software and the solid mechanics package.  In this case, to corroborate the experiments, a nearly incompressible Neo-Hookean material is assumed to model the Cheetah TPU hypereleastic mechanical behavior. The state variables of the problem are the displacements $\mathbf{u}$, the deformation gradient is therefore: $\mathbf{F} = \mathbf{I}+\nabla \mathbf{u}$. The free elastic energy $W$ corresponding to the Neo-Hookean hyperelastic material reads as:
\begin{equation}
W=\dfrac{1}{2}G(\mathbf{C}: \mathbf{I}-3)+\dfrac{1}{2}K (J-1)^2,
\end{equation}
where $G=9\cdot 10^7$ (Pa) is the shear modulus of TPU (fitting experimental results), $\mathbf{C}=\mathbf{F}^T\mathbf{F}$ is the Cauchy-Green tensor, $K = 33.44 \,G$ is the bulk modulus and $J = \textrm{det}\,\mathbf{F}$ is the volume deformation given by the determinant of the deformation gradient. Given the reference configuration $\mathbb{B}_d$ with boundary $\partial \mathbb{B}_d$ of unit normal $\mathbf{m}$, the problem is solved for the state variables satisfying the following balance equation:
\begin{equation}
\textrm{div}\,\mathbf{S}=0,
\end{equation}
where $\mathbf{S}=\dfrac{\partial W}{\partial \mathbf{E}}$ is the stress tensor associated to the Neo-Hookean hyperelastic free energy with $\mathbf{E}=\dfrac{1}{2}(\mathbf{F}^T\mathbf{F}-\mathbf{I})$ the elastic strain tensor. The balance equation is implemented in a weak formulation in the COMSOL software and supplemented by the boundary conditions on $\partial \mathbb{B}_d$. In particular, we control the displacements $\mathbf{u}$ at the boundaries of the TPU structure. To determine the stiffness of the reference structure, a shear deformation is directly applied, and the forces required to achieve that displacement are calculated in post-processing. This allows the shear stiffness to be estimated for State 0. 
Subsequently, the same shear deformation is applied to a second 3D model with the mirrored configuration of the curved beam to obtain the stiffness in State 1. The balance equation is solved using a stationary solver and an automatic Newton nonlinear method. 

\subsection{Fabrication process}

The monolithic soft clutch and the two-cell metamaterial specimens are designed in Rhinoceros 3D using the Grasshopper plugin. Both are fabricated by fused deposition modeling (FDM) on a Prusa XL multi-material printer, using Cheetah TPU filament (NinjaTek Inc., Shore hardness 95A) as the build material. For the soft clutch, a water-soluble butenediol vinyl alcohol copolymer (BVOH; Verbatim Inc.) is employed as the support material, while the two-cell metamaterials are printed directly without supports. Prior to printing, the TPU filament is dried in a Creality Space Plus dryer box at \SI{50}{\celsius} and \SI{25}{\percent} relative humidity. Printing is carried out with a layer height of \SI{0.2}{\milli\meter} and a \SI{90}{\percent} infill using standard grid patterns; external perimeters are deposited at a reduced nozzle speed of \SI{10}{\milli\meter\per\second} with an acceleration of \SI{10}{\milli\meter\per\second\squared}, while standard TPU parameters are applied to the internal regions. After fabrication, the clutch is immersed in water and sonicated for approximately \SI{4}{\hour}, ensuring complete dissolution of the BVOH and releasing the moving components without requiring manual assembly.

\subsection{Mechanical characterization}

The mechanical properties of the monolithic soft clutch and the two-cell metamaterial specimens are characterized using a ZwickRoell Z010 universal testing machine. For the clutch, the device is mounted by clamping its main holder structure in the top grip, while the central shaft is secured in the bottom grip for tensile loading. Before testing, the structure is pre-stretched by \SI{0.5}{\milli\meter} to ensure consistent initial conditions, after which the shaft is pulled at a constant rate of \SI{0.076}{\milli\meter\per\second}. This procedure is repeated for each of the four distinct clutch states (000, 100, 110, and 111) to determine the force–displacement relationship at different stiffness levels.

For the two-cell metamaterial specimens, a custom holder is 3D-printed in PLA to secure the structure during testing. The same overall setup and preparation are used, except that no pre-stretch is applied and the tensile loading is performed at a higher constant rate of \SI{0.16}{\meter\per\second}. Each specimen is first tested in State 1 followed by tests in State 0 under the same conditions.